# Mechanical properties of anionic asymmetric bilayers from atomistic simulations

Wenjuan Jiang 江文娟, Yi-Chun Lin 林宜君, Yun Lyna Luo 罗蕴*

Department of Pharmaceutical Sciences, College of Pharmacy, Western University of Health Sciences, Pomona, CA, USA.

**ABSTRACT**

Mechanotransduction, the biological response to mechanical stress, is often initiated by activation of mechanosensitive (MS) proteins upon mechanically induced deformations of the cell membrane. A current challenge to fully understand this process is to predict how lipid bilayers deform upon application of mechanical stress. In this context, it is now well established that anionic lipids influence the function of many proteins. Here, we test the hypothesize that anionic lipids could indirectly modulate MS proteins by alteration of the lipid bilayer mechanical properties. Using all-atom molecular dynamics simulations, we computed the bilayer bending rigidity ($K_C$), the area compressibility ($K_A$), and the surface shear viscosity ($\eta_m$) of 1-palmitoyl-2-oleoyl-sn-glycero-3-phosphocholine (PC) lipid bilayers containing or not phosphatidylserine (PS) or phosphatidylinositol bisphosphate (PIP$_2$) at physiological concentrations in the lower leaflet. Tensionless leaflets were first checked for each asymmetric bilayer model, and a formula for embedding an asymmetric channel in an asymmetric bilayer is proposed. Results from two different sized bilayers show consistently that the addition of 20% surface charge in the lower leaflet of PC bilayer by PIP$_2$ has minimal impact on its mechanical properties, while PS reduced the bilayer bending rigidity by 22%. As a comparison, supplementing the PIP$_2$-enriched PC membrane with 30% cholesterol, a known rigidifying steroid lipid, produces a significant increase in all three mechanical constants. Analysis of pairwise splay moduli suggests that the effect of anionic lipids on bilayer bending rigidity largely depends on the number of anionic lipid pairs formed during simulations. The potential implication of bilayer bending rigidity is discussed in the framework of mechanosensitive Piezo channels.

**SIGNIFICANCE STATEMENT**

Understanding how mechanical stimuli modulate the function of cell membranes and mechanosensitive ion channels remains a crucial challenge in biophysics. In this context, molecular dynamics simulations have recently highlighted the importance of lipid-protein interactions, especially those involving anionic lipids located in the inner leaflet of lipid bilayers, to membrane protein function. However, how the presence of anionic lipids directly impacts the mechanical properties of cell membrane has not been systematically investigated. This study investigates the three major mechanical properties of three asymmetric anionic bilayer models commonly used in membrane protein simulations, providing a solid foundation for rigorous computational studies of mechanosensitive channels and other membrane proteins.

**INTRODUCTION**

In our bodies, compressional forces are produced when cells differentiate and tumors grow. Shear forces are created when blood flow rubs against the vascular wall, and air flows into the lung. The amount of physical deformation produced by mechanical forces largely depends on the cell membrane's mechanical properties. Lipidomic and biophysical studies have identified a tremendous variety of lipid species organized in asymmetric and heterogeneous patterns in cytoplasmic (inner) and exoplasmic (outer) leaflets of plasma bilayer, with abundant anionic lipids in the inner leaflet[1]. Changes in expression levels of individual lipid species have been implicated in many diseases. Hence, determining the asymmetric membrane's influence on membrane protein function remains one of the key challenges for physiology and pathology.

Overcoming this challenge is especially important for studying mechanosensitive (MS) ion channels whose activations are very sensitive to membrane heterogeneity. The most studied MS channels so far are MscS, MscI, and MscL ion channels in prokaryotes. In eukaryotes, MS channels can be found in Piezo channels, the transient receptor potential (TRP) channels, two-pore-domain $K^+$ ($K2P$) channels, and OSCA/TMEM63 ion channels. MS channels retain their mechanosensitive property in artificial lipid bilayers (i.e., without additional cellular components) and hence sense mechanical forces transmitted from lipids in a process called the "force from lipids" (FFL) paradigm[2]. This paradigm can be classified into two groups: (1) lipid bilayer composition and (2) direct lipid-channel interactions. The former involves understanding the effect of the bilayer mechanical properties on the protein, whereas the latter focuses on specific local interactions between lipid molecules and protein residues. Both mechanisms have been explored using *in vitro* or cell-based experimental techniques, yet, the exact FFL mechanism remains unclear. This is because experimental measurements of interactions between lipids and proteins are mostly indirect and *in vivo* characterization of the lipid heterogeneity remains very challenging due to the high spatiotemporal resolution required to study the fluctuating nanoscale assemblies of lipids and proteins in living cells. MD simulations have been especially appealing in this respect because of their inherently high resolution and ability to quantify and dissect the global bilayer property and local lipid-protein interactions.

The role of the multicomponent asymmetric membrane has been increasingly demonstrated in MS channels and other membrane proteins[3]. Independent functional studies found that both membrane stiffness and negative-charged lipids are modulators of MS channels. For instance, bilayer thickness and stiffness have a differential effect on MscL and MscS, and anionic phosphatidylinositol increases the tension sensitivity of MscL [3c]. Piezo channels are not only sensitive to the membrane rigidity fine-tuned by cholesterol[4] and saturated fatty acids[5], but also to phosphatidylinositol bisphosphate (PIP$_2$) depletion [6], and cell surface flip-flop of anionic phosphatidylserine (PS) [7]. In asymmetric droplet bilayers, Piezo1 spontaneously opens with negatively charged lipids in the inner leaflet[8]. PIP$_2$ lipids also play a critical role in regulating inwardly rectifying K$^+$ channels [9], TRPV5 channels [10], and TRPM8 channels [11]. These observations brought up the important questions of whether those anionic lipids, predominantly present in the inner leaflet of the plasma membrane, modulate MS channel activity by direct binding and/or by changing the overall membrane mechanical properties?

Most atomistic membrane protein simulations over the past decades consist of a single-component bilayer made of phosphatidylcholine (PC), the major biological membrane component. In that case, the homogenous bilayer is simply necessary to maintain the stability of the membrane protein. As the role of specific lipids on protein functions being discovered increasingly, it becomes necessary to construct an asymmetric bilayer containing different lipid species due to the asymmetric nature of the plasma membrane. It is now well established that membrane rigidity is influenced by the size and charge of headgroups and the flexibility of fatty acid chains, such as the degree of unsaturation and *trans/cis* isomerization of double bonds. In general, charged lipids are believed to increase the overall $K_C$ if they induce a stronger repulsion between their lipid headgroups[12]. The importance of anionic lipids for membrane proteins, especially in MS channels, motivated us to investigate whether adding anionic lipids to PC bilayers alters intrinsic bilayer mechanical properties, which will, in turn, modulate channel functions.

Three important membrane mechanical properties are bilayer bending rigidity ($K_C$), bilayer area compressibility ($K_A$), and shear viscosity ($\eta$). $K_C$ and $K_A$ quantify the energetic cost associated with membrane bending and stretching/compressing its area. Together, they determine the membrane deformation free energy with and without membrane tension[13]. Shear viscosity ($\eta$) quantifies the bilayer shear deformation under different shear rates. Therefore, when simulating MS channel, either under equilibrium or under external stimuli (e.g., membrane stretching, cell poking, fluid shear stress), knowledge of these mechanical properties of the membrane model in which the channel is embedded is critical for rigorously assessing the outcome of the protein dynamics.

While the plasma membrane composition is rather complex, in all-atom MD simulations, heterogenous bilayer models are often limited to a few lipid types to ensure lateral distribution convergence. In this study, one symmetric palmitoyl-2-oleoyl-*sn*-glycero-3-phosphocholine (POPC) bilayer and two asymmetric POPC bilayers containing ~5% $PIP_2$ or ~20% POPS in the inner leaflet are investigated (named PC:PC, PC:PS and PC:$PIP_2$ bilayers, thereafter). 5% $PIP_2$ in the inner leaflet was chosen to reach the known percentage (~2.5%) in the plasma membrane. Since the dominant protonation state of $PIP_2$, suggested by NMR studies, carries -4 net charge[14], 5% PIP2 and 20% POPS (net charge -1e) allow us to compare the mechanical properties of two anionic bilayers with ~20% surface charge in the inner leaflet. For comparison, a ternary asymmetric bilayer composed of 30% cholesterol (about the typical sterol concentration in mammalian plasma membrane[15]) in both leaflets and 5% $PIP_2$ in the inner leaflet was also simulated (named PC:Chol:$PIP_2$). This ternary bilayer is known to be more rigid than PC:PC bilayer and is a useful bilayer model for simulating protein-cholesterol and protein-$PIP_2$ interactions.

We focused on testing the CHARMM36 all-atom lipid force field[16] that is frequently used in membrane protein simulations. Two bilayer sizes of ~200 and ~400 lipids were tested to ensure that the computed mechanical properties are converged within 500~900 nanoseconds (ns). First, leaflet tension was computed for each asymmetric bilayer to ensure there is no leaflet tension mismatch before calculating bending rigidity ($K_C$), bilayer area compressibility ($K_A$), and shear viscosity ($\eta$). All the bilayers simulated were fully hydrated and remain in the disordered lipid phase ($L_d$) that is most relevant to the membrane protein simulations at body temperature and allowed us to capture the bilayers in the Newtonian regime during the shear simulations.

## METHODS

### Bilayer setup and simulation protocols

All bilayers were built from CHARMM-GUI membrane builder[17]. Each system was simulated with NAMD2.11 and AMBER16 using the CHARMM36 lipid force fields and the CHARMM TIP3P water model[16]. Details about simulated systems are shown in **Table 1 and Table S1**. Each bilayer was solvated in 150 mM KCl solution. A non-bonded cut-off of 12.0 Å plus a 10.0 Å force switching range was employed. In the NVT ensemble, temperature control was done using Langevin thermostat with a gamma parameter (friction coefficient) of 1.0 $ps^{-1}$. The SHAKE algorithm was used to constrain bonds involving hydrogen. In the NPT ensemble, pressure regulation was achieved using a semi-isotropic Monte-Carlo barostat with a target pressure of the 1.0 bar and constant zero membrane surface tension. Simulation timestep was set to be 2.0 fs. Except for shear simulations (see below), production runs were conducted in the NPT ensemble. The atomic coordinates and velocities were saved every 10 ps for analysis. All simulations were conducted at 303.15 K (30 ℃).

**Leaflet tension in asymmetric bilayer model**

In CHARMM-GUI membrane builder, the numbers of lipids in each leaflet are estimated based on the individual area per lipid reported from symmetric bilayer studies. However, building an asymmetric bilayer by matching leaflet areas using area per lipid does not guarantee a tensionless bilayer within each leaflet. We hence first computed leaflet tension for each bilayer. The leaflet tension was computed from the lateral pressure profile of the bilayer following the method of Doktorova and Weinstein[18]. In brief, each bilayer was first aligned over the trajectory and centered on the average z position of all terminal methyl groups of the lipids at z = 0 Å. The box height along the z-axis was divided into 75 slabs with a thickness of ~1 Å. The three components of the pressure tensor were calculated in each slab with NAMD2.11. The total lateral pressure in each slab was the sum of Ewald and non-Ewald pressure contributions. The normal component of pressure in each slab $p_N$ was corrected to a constant value so that the integral of the whole pressure profile is zero. The tension in upper and lower leaflets were obtained by Eq. 1, where $p_L(z)$ is the lateral pressure in slab $z$.

$$T_{upper} = -\int_0^\infty (p_L(z) - p_N)dz \qquad T_{lower} = -\int_{-\infty}^0 (p_L(z) - p_N)dz \qquad \text{Eq .1}$$

**Leaflet area compressibility**

The area compressibility $K_A$ is an important mechanical property that quantifies the response of membrane area to tension. For symmetric bilayer with minimum undulations (the difference in projected areas and local areas is negligible), the area compressibility $K_A$ can be evaluated from the mean square fluctuation of the total area of the bilayer[19] or the probability distribution of the area change around the mean[20]. However, for asymmetric bilayer, the $K_A$ values for each bilayer leaflet are needed. Hence, we calculated $K_A$ based on local thermal fluctuations of the leaflet thickness[20]. In this approach, each leaflet is viewed as a collection of more than one parallel elastic blocks with the same average area but different instantaneous areas. The interleaflet coupling is shown to be equivalent to the variance of bilayer area $A$, ($\sigma^2(A)$). The local area fluctuation is then converted to the local thickness fluctuation assuming volume conservation[20]. The area compressibility of each leaflet $K_A^L$ can hence be obtained from the quadratic term in equation (2), where $t$ and $\delta t$ are the instantaneous local thickness of a leaflet and the deviation from equilibrium thickness, $a_0^L$ is the equilibrium local area of the leaflet, $C'$ is a constant.

$$-\frac{2k_B T}{a_0^L}\ln p(\frac{\delta t}{t}) = K_A^L\left(\frac{\delta t}{t}\right)^2 + C' \qquad \text{Eq.2}$$

It should be noted that the magnitude of leaflet thickness fluctuation depends on the location of the leaflet surface. For example, the rigid headgroup region may yield larger $K_A^L$ than the flexible fatty acid chain region. Doktorova et al. [20] proposed to use a correlation analysis of the lipid height fluctuation to determine a suitable location lying in between the rigid and flexible regions. For our non-Chol-containing leaflets, this surface is located at C8 or C9 atom of POPC, while for the 30% Chol-containing leaflets, this surface is located at C11 of POPC lipid (**Figure S2**). The probability distribution of the leaflet thickness using kernel density estimation and the quadratic fit using Eq (2) are shown in **Figure S3**. The $K_A$ of the bilayer is the harmonic mean of each leaflet. The error bars were calculated with the bootstrapping analysis as in ref[20].

The calculation of $K_A$ of the Chol-containing leaflet deserves special attention. It has been shown that at mol fraction of cholesterol less than 35%, compression between cholesterols is negligible, hence the $K_A^L$ can be estimated from the non-cholesterol components[17a, 20-21]. To calculate $K_A^L$ with Eq.2, the average area per non-Chol molecule, $a_{0(nonCh)}^L$, is needed and can be obtained from the relationship using Eq. 3[21b],

$$a_0^L = \chi_{nonCh} a_{0(nonCh)}^L + \chi_{Ch} a_{0(Ch)}^L \quad \text{Eq. 3}$$

where $\chi_{Ch}$ is Chol's mole fraction in the leaflet and $\chi_{nonCh=} = 1 - \chi_{Ch}$ is the mole fraction of the non-Chol components. Here, the area per Chol, $a^L_{0(Ch)}$, is approximated from Chol's average tilt angle as $a^L_{0(Ch)} = \frac{a'_{0(Ch)}}{\cos(\theta(\chi_{ch}))}$, in which $a'_{0(Ch)}$ of 0.38 nm$^2$ is the cross-sectional area of cholesterol with zero tilt defined in ref[21b]. The tilt angle $\theta$ of cholesterol is defined as the angle between the cholesterol director vector $\vec{n}$ (**Figure 1**) and bilayer normal $\vec{N}$. The bilayer normal is derived from the time-averaged lipid-water interface[22]. For each leaflet with cholesterol, we calculated the average cholesterol tilt angle $< \theta >$, defined in Eq. 4[23],

$$< \theta >= \int_0^{90} \theta P(\theta) d\theta \qquad \text{Eq. 4}$$

where $P(\theta)$ is the normalized probability density of cholesterol tilt angle in the PC:Chol:PIP$_2$ membrane.

**Bilayer bending rigidity**

Several methods have been developed to compute bilayer bending rigidity $K_C$. The most widely used one is the spectral analysis of the bilayer height thermal fluctuation or lipid tilt fluctuations during the MD simulations[24]. $K_C$ has also been calculated from forces exerted by a buckled membrane[25]. This method can be applied to complex heterogeneous bilayer and has been extensively tested on Martini coarse-grained force field[26]. Here, we use the local splaying fluctuation method, which has been tested on multicomponent bilayers using the same CHARMM36 all-atom force field[27] and can be in theory applied to the tensionless leaflets in an asymmetric bilayer. Following the method in ref[22], the lipid splay $S_i$ was computed from the covariant derivative of the vector field $\vec{n} - \vec{N}$ along one direction on the membrane interface. $\vec{N}$ is the membrane normal derived from the time-averaged lipid-water interface, and $\vec{n}$ is lipid director vector defined as a vector pointing from the lipid tail to its head (**Figure 1**). A cut-off distance of 10Å for phospholipid components were used to define neighboring lipid pairs. $K_C$ for each leaflet is obtained as the coefficient of the quadratic term (Eq. 5):

$$-\frac{2k_B T}{A_L} \ln P(S_i) = K_C(S_i)^2 + C \qquad \text{Eq.5}$$

where $A_L$ is the area per lipid in each leaflet (bilayer area divided by the total number of lipids per leaflet), and $C$ is a normalized constant. Each distribution of splays $P(S_i)$ was fitted to a gaussian to determine their mean $\mu$ and standard deviation $\sigma$. Error bars of $K_C$ are the standard deviation from quadratic fitting using for five fitting ranges in $[\mu - c\sigma, \mu + c\sigma]$, in which $c \in \{1, 1.25, 1.5, 1.75, 2\}$ [22]. For a multicomponent leaflet, the splay modulus $K_C^{ij}$ is calculated for all possible lipid pairs $i, j$. The aggregated leaflet $K_C$ is obtained by weighted harmonic mean using the number of $i, j$ pairs ($\varphi_{ij}$) and total pairs ($\varphi_{tot}$) as the weighting factor of each splay component (Eq.6).

$$\frac{1}{K_C} = \frac{1}{\varphi_{tot}} \sum_{i,j} \frac{\varphi_{ij}}{K_C^{ij}} \qquad \text{Eq.6}$$

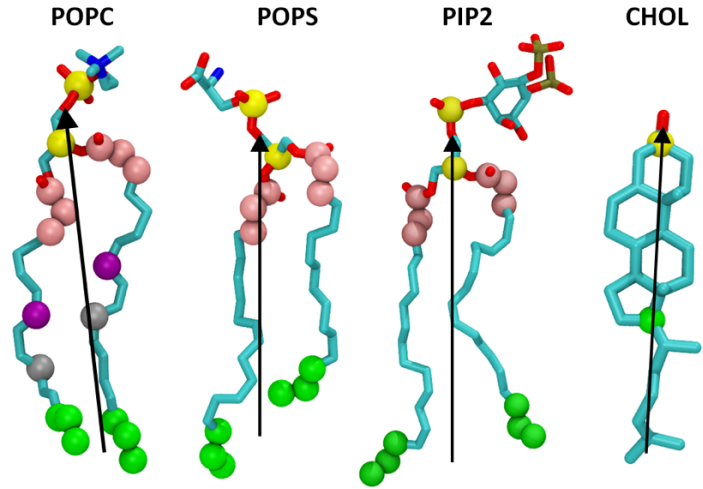

**Figure 1**. Schematic representation of POPC, POPS, PIP₂ and Chol lipids showing the definitions used for $K_C$ calculations. The $\vec{n}$ director vector (black arrow line) for each lipid pointing from lipid tails (center of mass of three terminal carbon atoms in green sphere in each tail) to the headgroup groups (mid-point between the headgroup P and C2 atoms in yellow sphere, for Chol lipid, lipid tail atom in green is C17 and headgroup atom in yellow is C3). Atoms used in lipid pairwise distance calculation are shown in pink sphere for POPC, POPS and PIP₂, while for Chol lipid, atom used is C3 in yellow sphere. The corresponding

selections for $K_C$ calculations using local splay fluctuation method are listed here: tails=[POPC: C216, C217, C218, C314, C315, C316; POPS: C216, C217, C218, C314, C315, C316; PIP2: P, C218, C219, C220, C316, C317, C318; CHOL: C17]; head groups=[POPC: P, C2, "POPS": P, C2; PIP2: P, C2; CHOL: C3]; distance selection =[POPC: C21, C22, C23, C31, C32, C33; POPS: C21, C22, C23, C31, C32, C33; PIP2: C21, C22, C23, C31, C32, C33; CHOL: C3]. C8 in purple and C11 in silver on POPC are used to define the surface for computing leaflet thickness fluctuation.

**Surface viscosity of lipid bilayers from nonequilibrium simulations**

Under external stimuli, such as shear flow, different viscous bilayers exert differential influence on the membrane protein conformational changes. Several equilibrium and nonequilibrium methods for computing shear viscosities of fluids from MD simulations have been discussed in ref[28]. Here, we calculate bilayer surface viscosity ($\eta_m$) from nonequilibrium deformation-based shearing simulations, which have been used previously for a coarse-grained force field[29] and recently for an all-atom force field[30]. In this method, the membrane surface viscosity was a 2D viscosity resisting perpendicular shear flow. For a membrane situated in the *xy*-plane, we set the shear flow in the *x*-direction with the gradient in the *y*-direction (**Figure 2**) using the box deformation function in GROMACS 2016.4 package. The membrane stress can be estimated from the total stress of the system $\sigma_{xy}^{TOT}$ minus the contribution from the solvent $\sigma_{xy}^{SOL}$ (Eq. 7):[29b]

$$\eta_m = \frac{\sigma_{xy}^{TOT} - \sigma_{xy}^{SOL}}{\dot{\varepsilon}} = \frac{-\langle P_{xy} \rangle L_z - \eta_w \dot{\varepsilon}(L_z - h)}{\dot{\varepsilon}} \qquad \text{Eq.7}$$

where $L_z$ is the height of the box, $\langle P_{xy} \rangle$ is the pressure tensor element, and $\eta_w$ is the viscosity of the solvent, $h$ is the membrane thickness. The average thickness values is calculated using the distance between two phosphate group density peaks. Shear rate $\dot{\varepsilon}$ is varied until the Newtonian regime is identified, where the surface viscosity does not depend on the shear rate. Each shear simulation was started from the final snapshot of the NPT run, and the average box size over the last 100 ns was used for NVT run with a series of strain rate of 0.02, 0.06, 0.1, 0.2 ns⁻¹ (except for PC:Chol:PIP₂ bilayer, 0.04 ns⁻¹ and 0.08 ns⁻¹ were also added additionally.). Simulation protocols were the same as above, except Nose-Hoover thermostat with a coupling constant of 1 ps was used for 303.15 K in GROMACS 2016.4 and the LINCS algorithm was used to constrain bonds involving hydrogen.

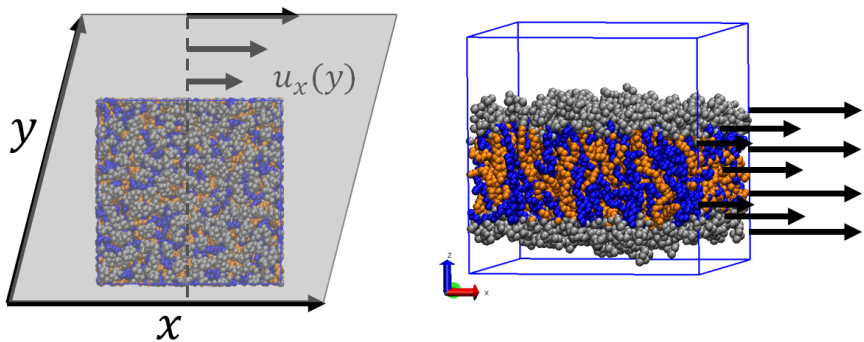

**Figure 2**. Shear deformation for symmetric POPC system. The left plot shows the shear flows along *x*-axis with gradient along *y*-axis, thus achieving the perpendicular shear flow. The right plot shows the flow gradient from the *xz*-plane view. The head groups of bilayers are shown in silver, one hydrophobic chain (C21-C218) of POPC bilayer is shown in blue, and the other chain (C31-C316) is shown in orange VDW mode (hydrogen atoms not shown).

## RESULTS AND DISCUSSION

### Tensionless leaflets in asymmetric bilayers

MD simulation of asymmetric bilayers using the periodic boundary condition requires that the surface areas of both lipid leaflets are equal. In other words, the relationship between leaflet area and leaflet tension depends on the compressibility modulus of the bilayer and not of each leaflet alone. Consequently, lipid deletion, insertion, and flip-flop in the bilayer may cause large and opposite leaflet tension, even when simulations are done under zero surface tension. In those cases, the validity of mechanistic insights, especially those of mechanosensitive channels, is compromised. Therefore, the comparison of asymmetric bilayer models requires that the tension in each leaflet equals zero.

The three anionic asymmetric bilayers were built based on the default area per lipid values (POPC 68.3, POPS 60.4, PIP$_2$ 67.4, cholesterol 40.0 Å$^2$) using the CHARMM-GUI membrane builder (**Table 1**). The tension for each leaflet was calculated from the lateral pressure profile of the bilayer (see Methods), which shows all the bilayers have leaflet tension magnitude less than 1 mN/m, similar to the ones from symmetric POPC bilayer (**Table 1**). Thus, the current ratio of upper/lower leaflet composition satisfies the zero-leaflet tension condition so that the bilayer can be considered tensionless. There is hence no need to adjust further those binary bilayer compositions used here. However, it is expected that, as the bilayer composition becomes more complex, leaflet tension mismatch becomes more likely to occur. For example, Doktorova and Weinstein[18] showed that in a bilayer built based on default area per lipid containing DPPC/Cholesterol in the top leaflet and SOPC in the bottom leaflet, the leaflet tension magnitude was as large as $5.7 \pm 1.0$ mN/m. In that case, the number of lipids in each leaflet must be adjusted to minimize leaflet tension.

After ensuring zero-leaflet tension in all simulated bilayer models, we can discuss the impact of anionic lipids on the lateral pressure of the POPC bilayer. The *x*-axis of the lateral profile indicates the distance from the bilayer center along membrane normal, with the lower leaflet at the negative side and the upper leaflet at the positive side (**Figure 3**). Except for the Chol-containing bilayer, the lateral pressure distribution of PC:PS and PC:PIP$_2$ bilayers is quite similar to the symmetric PC bilayer since PC is the dominant lipid component in both leaflets. The common features are the negative pressure at a distance $\pm 18$ Å (with respect to the center of the bilayer) mainly due to the hydrogen bonding between the PC glycerol linker region, and the positive pressure at a distance $\pm 22$ Å, mainly due to charge-charge repulsion mediated by the PC headgroups (see lipid charge density and mass density plots in **Figure S1**). However, in the lower leaflet, the width of this positive peak (-20<z<-30 Å) is reduced by the presence of anionic headgroups in both PC:PS and PC:PIP$_2$ bilayers. **Figure S1** shows that PS and PIP$_2$ both reduced the repulsive electrostatic interactions between cationic choline groups (z ~ -24 Å) among PC lipids. The lateral pressure profile of PC:Chol:PIP$_2$ bilayer is dominated by the typical features of cholesterol-containing PC bilayers that have been thoroughly discussed in ref[31].

Besides ensuring zero-leaflet tension in an asymmetric bilayer model, the ratio of total number of lipids in the upper and lower leaflets ($\frac{N_{upper}}{N_{lower}}$) obtained from a tensionless bilayer is also important for constructing membrane protein models. When embedding an asymmetric channel in an asymmetric bilayer with a total lateral area of $A_{xy}$, the amount of lipids in each leaflet should be further adjusted by the area occupied by the protein in the upper and lower leaflet ($A_{prot}^{upper}$ and $A_{prot}^{lower}$), so that the desired lipid ratio is $\frac{(A_{xy} - A_{prot}^{upper})}{(A_{xy} - A_{prot}^{lower})} \times \frac{N_{upper}}{N_{lower}}$. Although this formula does not factor in the change in area per lipid due to specific protein-lipid interactions, it should be used to alleviate, at least, the artifact introduced by leaflet tension mismatch from an asymmetric bilayer model.

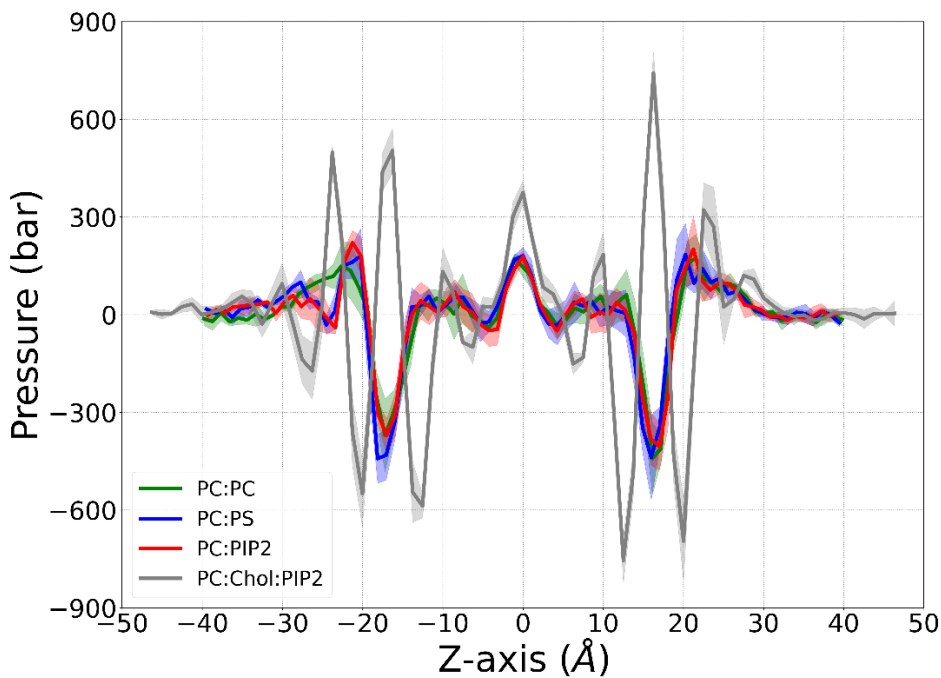

**Figure 3**. Lateral pressure profiles of symmetric and asymmetric bilayers. The lipid composition and leaflet tension computed from each bilayer are reported in **Table 1**. The lateral pressure profile for each system was averaged over 10,000 frames (last 100 ns). The standard deviations (shaded regions) were calculated from four blocks of 25 ns from the last 100ns for each system.

**Table 1. Mechanical properties of three anionic asymmetric bilayer models and a symmetric POPC bilayer using CHARMM36 force field**

| Simulated bilayers | Symmetric PC:PC | Asymmetric PC:PS | Asymmetric PC:PIP$_2$ | Asymmetric PC:Chol:PIP$_2$ |
|---|---|---|---|---|
| Upper leaflet (POPC) $N_{upper}$ | 100 | 100 | 104 | 105(75:30) |
| Lower leaflet $N_{lower}$ | 100 | 102(85:17) | 105(100:5) | 105(70:30:5) |
| Water/lipid ratio | 45.2 | 44.7 | 44.5 | 39.8 |

| Simulation time (ns)[a] | 500 | 500 | 500 | 400 |
|---|---|---|---|---|
| Leaflet tension $T_{upper}$ (mN/m) | 0.73 (±0.9) | 0.78 (±0.4) | 0.77 (±1.5) | 0.55 (±1.2) |
| Leaflet tension $T_{lower}$ (mN/m) | -0.73 (±0.9) | -0.78(±0.4) | -0.77 (±1.5) | -0.55 (±1.2) |
| Upper leaflet compressibility $K_A^L$ (dyn/cm) | 263 (±39) | 236 (±41) | 232 (±35) | 539 (±72) |
| Lower leaflet compressibility $K_A^L$ (dyn/cm) | 246 (±40) | 242 (±42) | 260 (±39) | 593 (±89) |
| **Bilayer $K_A$ (dyn/cm; harmonic mean)** | **254 (±29)** | **239 (±30)** | **244 (±26)** | **564 (±57)** |
| Upper leaflet bending modulus $K_C$ ($k_B$T) | 11.3 (±0.8) | 9.6 (±0.8) | 9.4 (±0.7) | 29.2 (±0.8) |
| Lower leaflet bending modulus $K_C$ ($k_B$T) | 10.8 (±0.7) | 7.6 (±0.2) | 12.1 (±0.5) | 28.6 (±1.2) |
| Upper leaflet $A_L$ (Å²)[b] | 64.3 (±0.8) | 65.1 (±1.0) | 64.9 (±0.7) | 48.1 (±0.7) |
| Lower leaflet $A_L$ (Å²) | 64.3 (±0.8) | 63.8 (±1.0) | 64.3 (±0.7) | 48.1 (±0.7) |
| **Bilayer bending Rigidity $K_C$ ($k_B$T)** | **22.0 (±1.5)** | **17.2 (±1.0)** | **21.5 (±1.2)** | **57.8 (±2.0)** |
| Bilayer Thickness $h$ (Å)[c] | 39.3 (±0.6) | 39.2 (±0.6) | 39.2 (±0.5) | 44.9 (±0.1) |
| **Surface shear viscosity $\eta_m$ (10⁻¹¹ Pa·m·s)** | **18.3 (±0.6)** | **19.8 (±0.9)** | **18.6 (±0.7)** | **39.5 (±0.3)** |

[a.] All simulations were conducted using semi-isotropic pressure coupling at 303.15K. [b.] Leaflet areas were calculated using the $xy$-dimension of bilayer divided by the total number of lipids in each leaflet. [c.] Bilayer thickness was calculated using the distance between two phosphate group density peaks. Error bars for leaflet tensions, leaflet area, and membrane thickness are standard deviations from the last 4 blocks of 25 ns. $K_A$ and $K_C$ were averaged from last 200 ns (time convergence are sown in **Figure S4 and S5**). Error bars of $K_A$, $K_C$, and $\eta_m$ were described in each Methods section.

## Bilayer area compressibility and bending rigidity

Changes in the lipid environment surrounding a membrane protein are likely to change its conformational free energy landscape, potentially changing the statistical distribution of individual proteins between functional states. It is thus not surprising that mechanical membrane deformations have profound effects on lipid-protein interactions and protein function. $K_A$ is a measure of the stiffness of the membrane in the lateral dimension. $K_C$ is a measure of the stiffness of the membrane normal to the membrane plane, i.e., the energetic cost of out-of-plane deformation. The total membrane deformation free energy hence depends on the area compressibility ($K_A$) and bending rigidity ($K_C$), and membrane tension[13].

We first tested our computational approach (see Methods) on the symmetric PC:PC bilayer. Using the last 200 ns of 500 ns simulations at 303.15 K, the symmetric PC:PC bilayer has a $K_A$ of $254 \pm 29$ dyn/cm (**Table 1**), in good agreement with previous simulations at 303 K (206 dyn/cm[20] and 240-280 dyn/cm[24]) and with experimental data (180-330 dyn/cm)[32]. Likewise, $K_C$ of the PC:PC bilayer is $22.0 \pm 1.5$ $k_B$T, close to the previous results from simulations (25.3 $k_B$T at 298K[27a] and 31.7 $k_B$T at 303K [24]) and experiments (8.5-8.6 x 10^-20 J =20.3-20.6 $k_B$T at 303K)[33].

Compared with a pure PC:PC bilayer, the addition of 20% negative surface charge in the lower leaflet by PS or PIP$_2$ has no significant effect on $K_A$ (only 6% and 4% decrease in the average value with error bars largely overlapping) (**Table 1** and **Figure S4**). For $K_C$, PIP$_2$ shows no significant effect, but PS imparted a 22% decrease in $K_C$, from $22.0 \pm 1.5$ $k_B$T to $17.2 \pm 1.0$ $k_B$T (**Table 1 and Figure S5**). In comparison, 30% cholesterol in the PC:PIP$_2$ bilayer increased $K_A$ by 122%, and $K_C$ 163%. Interestingly, a previous simulation study of a symmetric PC/PS bilayer using the same force field showed that 30% PS in both leaflets increased $K_A$ about 1.4 fold [20] and increased $K_C$ from 25.3 to 30.7 $k_B$T at 20°C[27a]. Thus, in our study, adding ~20% PS in the inner leaflet to mimic the charge asymmetry in the plasma membrane yields an effect on the $K_C$ opposite of the symmetric PC/PS bilayer with 30% PS in both leaflets. It should not be surprised that the effect of anionic lipids on bilayer rigidity is highly dependent on the lipid composition, hence cannot be generalized by the assumption that charged lipids will make bilayer more rigid through electrostatic repulsion. Below we show that the overall effect of anionic lipids on $K_C$ largely depends on the number of anionic lipid pairs.

Since the overall leaflet $K_C$ is obtained by weighted harmonic mean of each splay component (Eq.6), we can compare the splay modulus $K_c^{ij}$ for each lipid-pair type (**Table 2**). Within the PC-PIP$_2$ bilayer, $K_c^{ij}$ of PIP$_2$-PIP$_2$ pairs is 10 $k_B$T higher than PC-PC pairs (22.2 vs 12.0 $k_B$T) due to the highly charged nature of PIP$_2$ lipids. But the total $K_C$ is the harmonic mean of individual $K_c^{ij}$ weighted by number of pairs. The small number of PIP$_2$-PIP$_2$ pairs ($\varphi_{ij}$ 0.8 out of 382 total pairs per frame) resulted in a negligible contribution of PIP$_2$-PIP$_2$ pairs in the total $K_C$. In contrast, the $K_c^{ij}$ of PS-PS and PC-PS pairs are only 0.3-1.0 $k_B$T higher than PC-PC pairs. Thus, a much larger mole fraction of PS is needed to significantly increase the total $K_C$, as shown in the previous symmetric PC:PS (70/30) bilayer. Here, the presence of 20% PS in the lower leaflet decreases the PC-PC $K_C^{ij}$ (7.3 $k_B$T in PC:PS vs 10.8 $k_B$T in PC:PC lower leaflet). Consequently, the total $K_C$ of PC:PS bilayer is decreased. The rigidifying effect of cholesterol is reflected by a ~3 folds increase in PC-PC and PC-PIP$_2$ pairwise $K_C^{ij}$, compared with the PC-PIP$_2$ bilayer without cholesterol. The convergence of the number of pairs $\varphi_{ij}$ and weighted splay component $\frac{\varphi_{ij}}{\varphi_{tot} K_C^{ij}}$ in each lower leaflet are shown in **Figure 4**.

**Table 2. Pairwise lower leaflet $K_C^{ij}$ ($k_B$T) and $\varphi_{ij}$ for asymmetric bilayers.**

| Bilayer System | $\varphi_{tot}$ | PC-PC | | PC-PS/PIP$_2$ | | PS-PS or PIP$_2$-PIP$_2$ | | Chol-Chol | | Chol-PC | | Chol-PIP$_2$ | |
|---|---|---|---|---|---|---|---|---|---|---|---|---|---|
| | | $\varphi_{ij}$ | $K_C^{ij}$ | $\varphi_{ij}$ | $K_C^{ij}$ | $\varphi_{ij}$ | $K_C^{ij}$ | $\varphi_{ij}$ | $K_C^{ij}$ | $\varphi_{ij}$ | $K_C^{ij}$ | $\varphi_{ij}$ | $K_C^{ij}$ |
| PC:PS | 237.9 | 159.7 | **7.3** | 69.9 | **8.4** | 8.3 | **7.6** | | | | | | |
| PC:PIP2 | 382.1 | 347.0 | **12.0** | 34.3 | **12.1** | 0.8 | **22.2** | | | | | | |
| PC:Chol:PIP2 | 565.4 | 238.6 | **33.4** | 35.5 | **37.4** | 0.4 | **15.0** | 45.6 | **38.6** | 229.8 | **23.4** | 15.4 | **28.2** |

[*]$K_C^{ij}$ is splay modulus for lipid pairs $i, j$. $\varphi_{ij}$ is number of $ij$ pairs per frame and $\varphi_{tot}$ is number of total lipid pairs per frame (see Eq. 6). All values were computed from the last 1000 frames (100 ns). The convergence of $\varphi_{ij}$ and normalized $K_C^{ij}$ are shown in **Figure 4**.

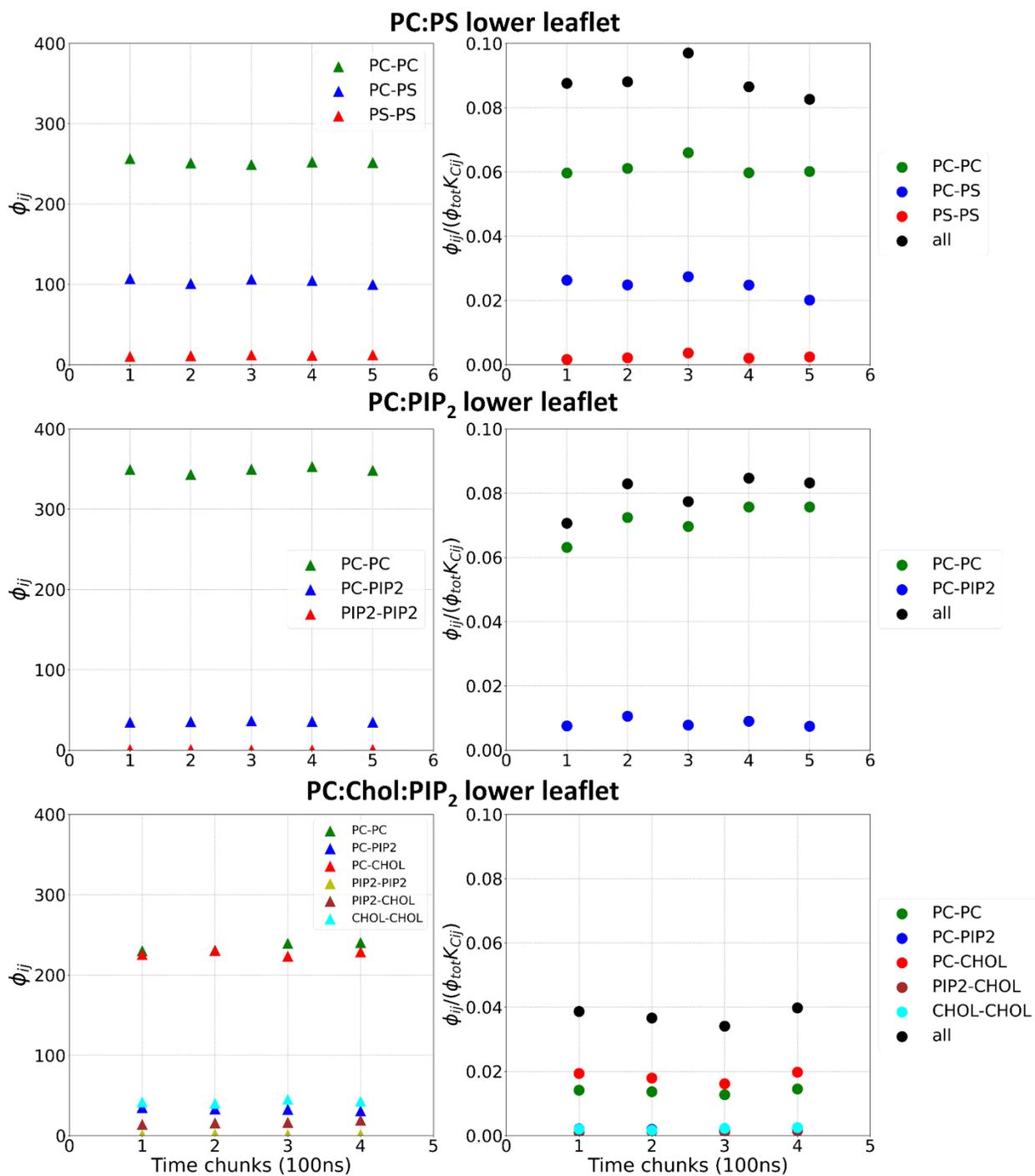

**Figure 4.** The convergency of the number of pairwise $i, j$ pairs $\varphi_{ij}$ (left panels) and weighted splay component $\frac{\varphi_{ij}}{\varphi_{tot} K_C^{ij}}$ (right panels) for the lower leaflet in each bilayer (see Eq. 6). Each time chunk is computed using 1000 frames (100 ns). The weighted splays of PIP$_2$-PIP$_2$ pairs are less than 0.001, thus not shown here.

The methods used here to compute $K_A$ and $K_C$ for asymmetric bilayers depend only on the local leaflet thickness fluctuations (Eq. 2) and lipid pairwise splay modulus (Eq. 5 and 6). The local nature of the analysis is expected to produce the results less sensitive to bilayer size. To check this hypothesis, we constructed three large-size bilayers by doubling the number of lipids while keeping the ratio of the lipids from the zero-leaflet tension bilayers (**Table S1**). The comparison and time convergence of the $K_A$ and $K_C$ for both small bilayers and large bilayers are reported in **Figure S4-S6**. Comparing the results for small bilayers (~200 lipids) in **Table 1** with the large bilayers (~400 lipids) in **Table S1**, we found that the $K_C$ results are nearly identical, while $K_A$ in the large bilayers are 5~16% lower than the small bilayers, although the error bars are overlapping. The reduced $K_A$ in our large bilayers is in fact consistent with the previous simulations reporting that the $K_A$ of a larger POPC bilayer with 416 lipids was underestimated due to larger undulation (the bilayer normal is not the same throughout the surface)[20]. This is because to calculate the leaflet thickness at a grid point, the height at each grid point is obtained by performing interpolation on the z-coordinates of the corresponding atoms. For a flat leaflet, the interpolation can be done over the whole leaflet surface. We found that constraining the radius of interpolation to 40 Å (slightly larger than the average bilayer thickness) alleviates the underestimation to a large extent, but not completely.

**Shear viscosity of bilayer membrane**

Mechanosensitive channels, such as Piezo[6], K$_2$P[34], and TRP[35], have been reported to sense fluid shear stress. Studies have suggested that shear stress tends to increase the lateral diffusion of individual lipids and reduce lipid order. Same as area compressibility and bending rigidity, different membrane compositions are likely to affect shear viscosity. A recent study of CHARMM36 lipid force field shows that the shear viscosity of DOPC bilayer at 25℃ is about twice that of DPPC at 50℃[30]. Here, we aim to investigate whether adding anionic lipids in the lower leaflet of a PC bilayer has any effect on shear viscosity at body temperature. Such information is necessary for future simulation studies of membrane proteins under shear stress.

A series of strain rates of 0.02, 0.06, 0.1, 0.2 ns$^{-1}$ (for PC:Chol:PIP$_2$ bilayer, 0.04 ns$^{-1}$ and 0.08 ns$^{-1}$ were added additionally) was explored for each bilayer system with three replicas of 50 ns were conducted for each strain rate. For all three non-Chol containing bilayers, the Newtonian region is reached at a strain rate between 0.06 ns$^{-1}$ and 0.1 ns$^{-1}$. For PC:Chol:PIP$_2$ bilayer, the Newtonian region is at between 0.08 ns$^{-1}$ and 0.1 ns$^{-1}$. The inverse-variance weighted averages from the Newtonian region yield similar surface viscosities for PC:PC, PC:PS, and PC:PIP$_2$, and 2.1 fold increase in the viscosity of PC:Chol:PIP$_2$, compared with PC:PIP$_2$ (**Table 1 and Figure 5**).

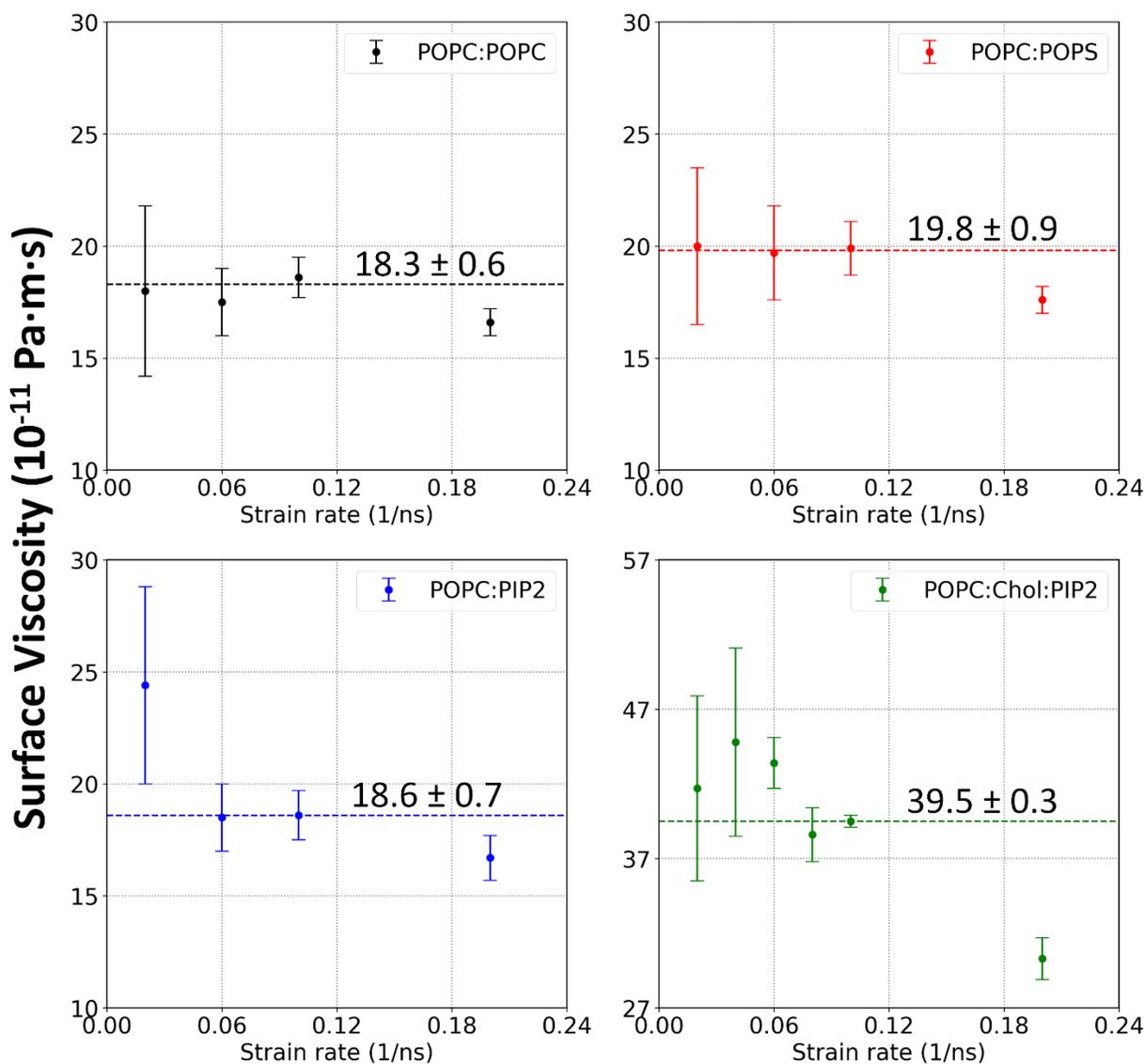

**Figure 5.** Surface viscosity as a function of the strain rate for three C36 bilayer systems. All systems were equilibrated independently to zero surface tension before shearing. Dotted lines show the result of taking inverse-variance weighted averages for strain rates between 0.06 ns$^{-1}$ and 0.1

ns$^{-1}$. Except for PC:Chol:PIP$_2$ bilayer, the result is between 0.08 ns$^{-1}$ and 0.1 ns$^{-1}$. Each data point is an average of three trails at the given strain rate.

**Implication for mechanosensitive channels**

Lipid-mediated modulation of protein function can be polymodal. Some modulatory effects are short-range and direct (e.g., protein-lipid binding), others are short-range but indirect (e.g., hydrophobic mismatch between protein and lipids), or even long-range (e.g., protein-induced membrane bending). The energetic cost of mechanical deformation of membrane, involving hydrophobic mismatch or bending, has been found to not only regulate protein function at the single-protein level, but also to influence protein localization, dimerization, and cooperative interactions between proteins[3a, 3c, 36]. For example, the curved structure of the mechanosensitive Piezo1 channel deforms the surrounding membrane into an inverted dome shape that extends beyond the protein boundary, called Piezo's footprint[37]. The decay length of the Piezo footprint has been estimated to be $\sqrt{K_C/\gamma}$, in which $\gamma$ is the membrane tension[37]. This footprint flattening has been shown to flatten Piezo1, enabling its pore to open[38]. The free energy change associated with Piezo1 channel opening can thus be approximated using Eq. (8):

$$\Delta G = \Delta G_{prot} - \Delta G_{memb} - \gamma \Delta A \qquad \text{Eq (8)}$$

in which $\Delta G_{prot}$ is the free energy of protein conformational change in the absence of membrane tension and $\Delta G_{memb}$ is the free energy cost associated with membrane bending[37]. $\Delta A$ is the relative change in the total projected area. When channels are near each other, an additional membrane deformation energy term due to the footprint overlap has been suggested to link to the inter-channel cooperativity in Piezo clusters [36c, 38]. According to the classic Helfrich's expression, the bending free energy of a bilayer normal to z-axis depends on the membrane curvature and bending rigidity (Eq. 9).

$$\Delta G_{memb} = \int \frac{1}{2} K_C \left(c_x + c_y\right)^2 dx dy \qquad \text{Eq(9)}$$

where $c_x$ and $c_y$ are the local values of the membrane curvature along the $x$ and $y$ axes defining the bilayer plane. We show that the symmetric PC and two asymmetric anionic bilayer models have their $K_C$ similar to ~20 $k_B$T documented for a typical plasma membrane[39]. Thus, the size or decay length of the Piezo footprint simulated in those bilayer models is likely to represent the shape of the Piezo1 footprint in cell membranes. **Table 1** shows that the magnitude of $\gamma$ fluctuation is on the order of 1 mN/m (0.24 $k_B$T/nm$^2$ at T = 303.15) during the tensionless bilayer simulation. The Piezo1 membrane footprint will thus extend ~10 nm beyond the protein boundary when simulated in PC:PC, PC:PS or PC: PIP$_2$ bilayers. However, the asymmetric PC:Chol:PIP$_2$ bilayer has a bending rigidity >30 $k_B$T more rigid than a plasma membrane. Hence, when simulating Piezo1 protein in PC:Chol:PIP$_2$ bilayer, one would expect a larger membrane footprint with decay length ~15 nm. With the larger $K_C$ and larger total deformation area in the integrand of Eq(9), $\Delta G_{memb}$ could be increased by hundreds of $k_B$T under small membrane tension[37]. Although variations in lipid composition and/or redistribution between two leaflets (e.g., cholesterol flip-flop) might occur to compensate the energetic penalty of membrane deformation to some extent while maintaining cell homeostasis, such reorganization will not occur within the current all-atom simulation timescale. Therefore, caution is needed to interpret physiological relevant results if the protein is simulated in a bilayer that has rather different mechanical properties than those in the plasma membrane.

## CONCLUSION

As more computational studies move away from homogenous PC bilayers and towards more complex asymmetric bilayers, the different mechanical properties of the bilayers could play an important role in shaping the protein conformational ensemble. Owing to the importance of anionic lipids on protein functions, PS and PIP$_2$ lipids are often included in the inner leaflet of the PC bilayer with or without cholesterols during the membrane protein MD simulations. Using all-atom MD simulations of three tensionless asymmetric bilayers, we showed that including anionic lipids PS and PIP$_2$ at the physiological concentration (about 20% surface charge in the inner leaflet) has a minimum impact on the bilayer area compressibility and surface shear viscosity. For bending rigidity, while PIP$_2$ has no effect, PS imparted a moderate (22%) decrease. In comparison, the asymmetric bilayer including PIP$_2$ and 30% cholesterols resulted larger than 2-fold increase in all three mechanical constants. The effect of charged lipids on bending rigidity is further analyzed using pairwise splay modulus. We found that the overall effect of anionic lipids on $K_C$ largely depends on the number of anionic lipid pairs, thus should not be generalized by the assumption that charged lipids will make bilayer more rigid due to charge-charge repulsion. In addition, we discussed in a semiquantitative way how the bilayer mechanical properties will impact the outcome of the membrane protein simulations using Piezo mechanosensitive channel as an example.

It remains to be seen whether the exact values of the bilayer properties calculated here match the "reality" when experimental results are available. The computed mechanical properties are highly sensitive to the atomistic force fields, which are subject to continuous improvement. For example, it is known that the current C36 lipid force field lacks the long-range Lennard-Jones interaction. The new CHARMM36 lipid force field with an explicit treatment of long-range dispersion[40] would be expected to increase the membrane surface shear viscosity. Although the mechanical properties reported from current atomistic bilayer models do not represent the ones from the plasma membrane, those binary and ternary membrane models are frequently used in MD simulations of ion channels and transporters. Caution is needed to interpret the influence of lipids on proteins since embedding membrane proteins will also change membrane topology (bending and thinning), and the annular lipids do not behave in the same way as bulk lipids. Such effect, although not investigated here, could alter protein function. Quantifying the relative mechanical properties between different bilayer models is perhaps the first critical step to study the force-from-lipids paradigm, which includes the overall mechanical force from bilayer and/or specific protein-lipid interactions.

**Data availability**

The code and data presented in this study are openly available in https://github.com/LynaLuo-Lab/memb-mechanical-properties.

**ACKNOWLEDGMENTS**

Authors wish to thank Dr. Milka Doktorova for her valuable help and discussion on the calculation of $K_A$ ad $K_C$, and Dr. Jerome J. Lacroix for discussion and proofreading. This work was supported by NIH Grants R01-GM130834. Computational resources were provided via the Extreme Science and Engineering Discovery Environment (XSEDE) allocation TG-MCB160119. The XSEDE program is supported by NSF grant number ACI-154862.

# SUPPORTING INFORMATION

# Mechanical properties of anionic asymmetric bilayers from atomistic simulations

Wenjuan Jiang, Yi-Chun Lin, Yun Lyna Luo*

Department of Pharmaceutical Sciences, College of Pharmacy, Western University of Health Sciences, Pomona, CA, USA.

**Table S1. Mechanical properties of large-size bilayer system using CHARMM36 force field**

| Simulated large bilayers | Symmetric PC:PC | Asymmetric PC:PS | Asymmetric PC:PIP$_2$ |
|---|---|---|---|
| Upper leaflet (POPC) $N_{upper}$ | 200 | 200 | 208 |
| Lower leaflet $N_{lower}$ | 200 | 204(170:34) | 210(200:10) |
| Water/lipid ratio | 45.1 | 44.5 | 44.5 |
| Simulation time (ns)[a] | 900 | 900 | 900 |
| Upper leaflet compressibility $K_A^L$ (dyn/cm) | 234 (±41) | 228 (±36) | 194 (±30) |
| Lower leaflet compressibility $K_A^L$ (dyn/cm) | 195 (±37) | 225 (±43) | 222 (±38) |
| **Bilayer $K_A$ (dyn/cm; harmonic mean)** | **213 (±28)** | **225 (±30)** | **206 (±24)** |
| Upper leaflet bending modulus $K_C$ ($k_B$T) | 12.1 (±0.8) | 9.2 (±0.7) | 8.8 (±0.5) |
| Lower leaflet bending modulus $K_C$ ($k_B$T) | 10.0 (±0.8) | 8.2 (±0.2) | 12.3 (±0.4) |
| Upper leaflet $A_L$ (Å$^2$)[b] | 65.0 (±0.8) | 64.7 (±0.8) | 65.0 (±0.7) |
| Lower leaflet $A_L$ (Å$^2$) | 65.0 (±0.8) | 63.4 (±0.8) | 64.4 (±0.7) |
| **Bilayer bending Rigidity $K_C$ ($k_B$T)** | **22.1 (±1.6)** | **17.4 (±0.9)** | **21.1 (±0.9)** |
| Bilayer Thickness $h$ (Å)[c] | 38.8 (±0.5) | 39.3 (±0.4) | 38.9 (±0.3) |

[a]All simulations were conducted using semi-isotropic pressure coupling at 303.15K. [b]Leaflet areas were calculated using the $xy$-dimension of bilayer divided by the total number of lipids in each leaflet. [c]Bilayer thickness was calculated using the distance between two phosphate group density peaks. Standard deviations for leaflet area and membrane thickness were calculated using four blocks of 25 ns over the last 100 ns. $K_A$ and $K_C$ values were averaged from last 400 ns. $K_A$ and $K_C$ convergence are shown in **Figure S4 and S6**. Error bars of $K_A$ and $K_C$ were described in each Methods section.

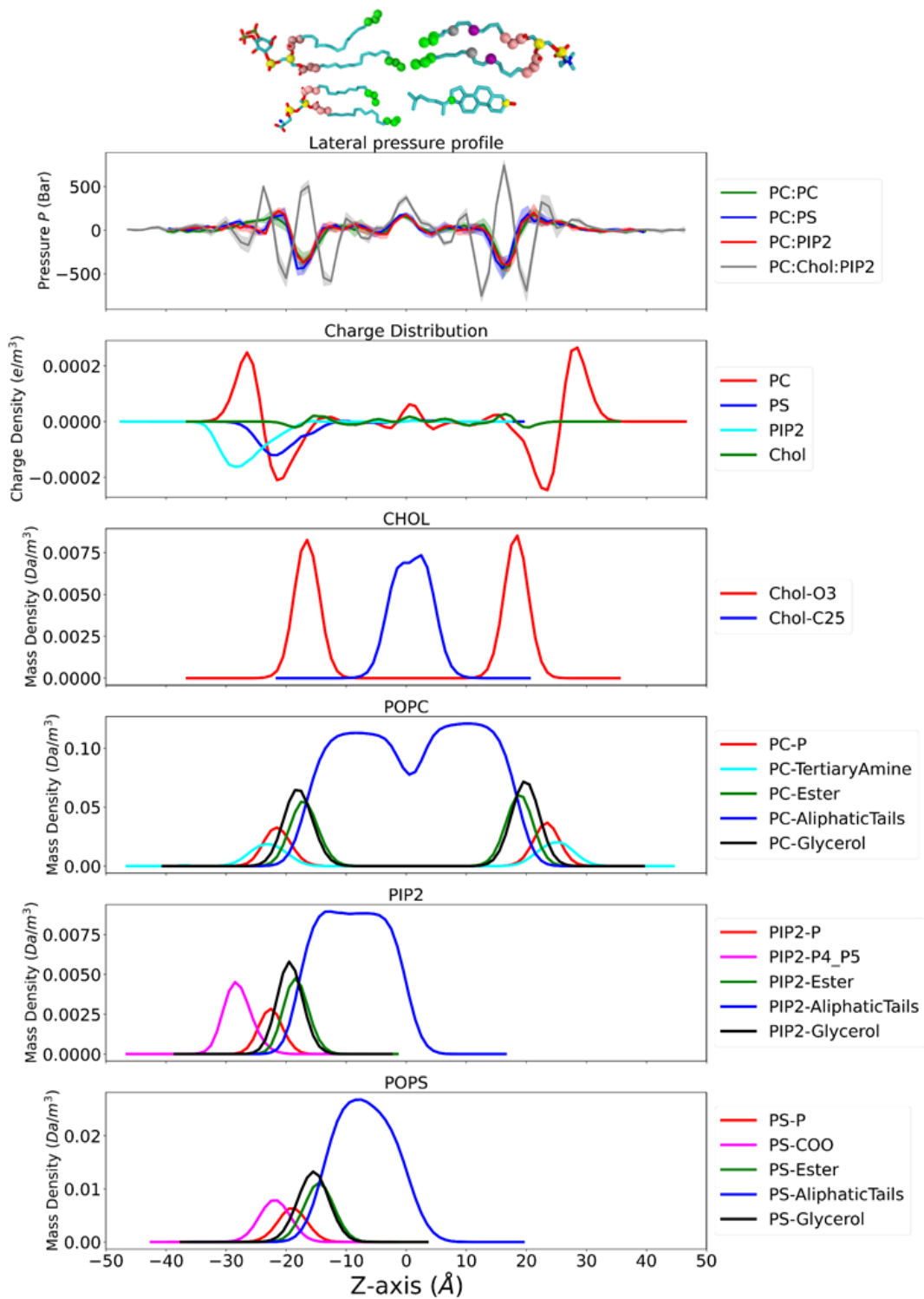

**Figure S1**. Lateral pressure profiles of bilayers with standard deviations (shaded) from 4x25 ns from the last 100ns (top), charge density distributions (the 2nd row), group-based mass density distributions (3rd to 6th rows) for different bilayer systems.

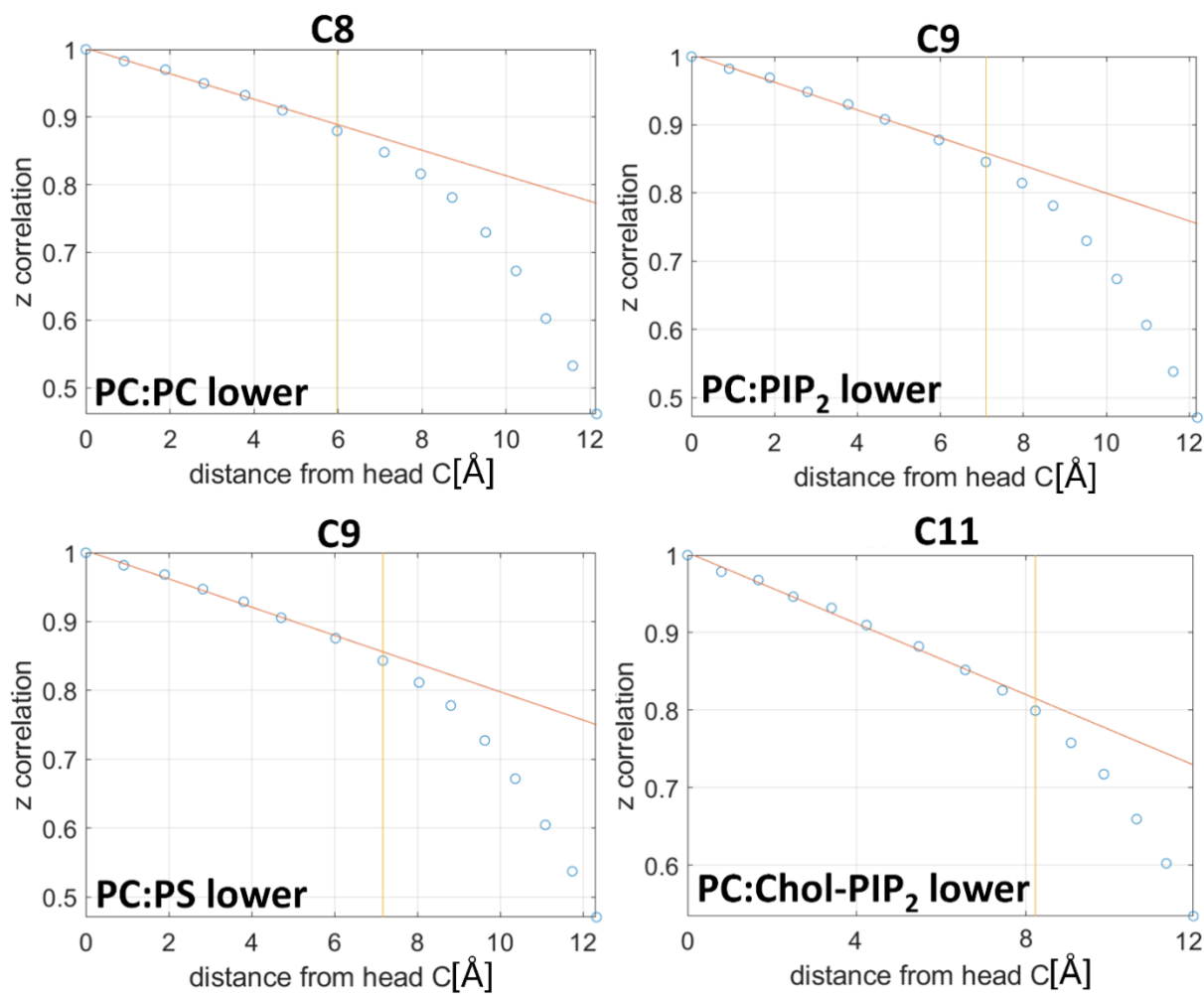

**Figure S2**. Correlation analysis of the lipid height fluctuation between acyl-chain carbons. For each system, $x$-axis stands for the distance between each acyl-chain carbon and the reference carbon C2; $y$-axis stands for the correlation values. The red line shows the best-fit line. The yellow line marks the first point outside of the linear regime that is used to extract the leaflet thickness fluctuations: C8 for PC:PC, C9 for PC:PS and PC:PIP$_2$, C11 for PC:Chol-PIP$_2$.

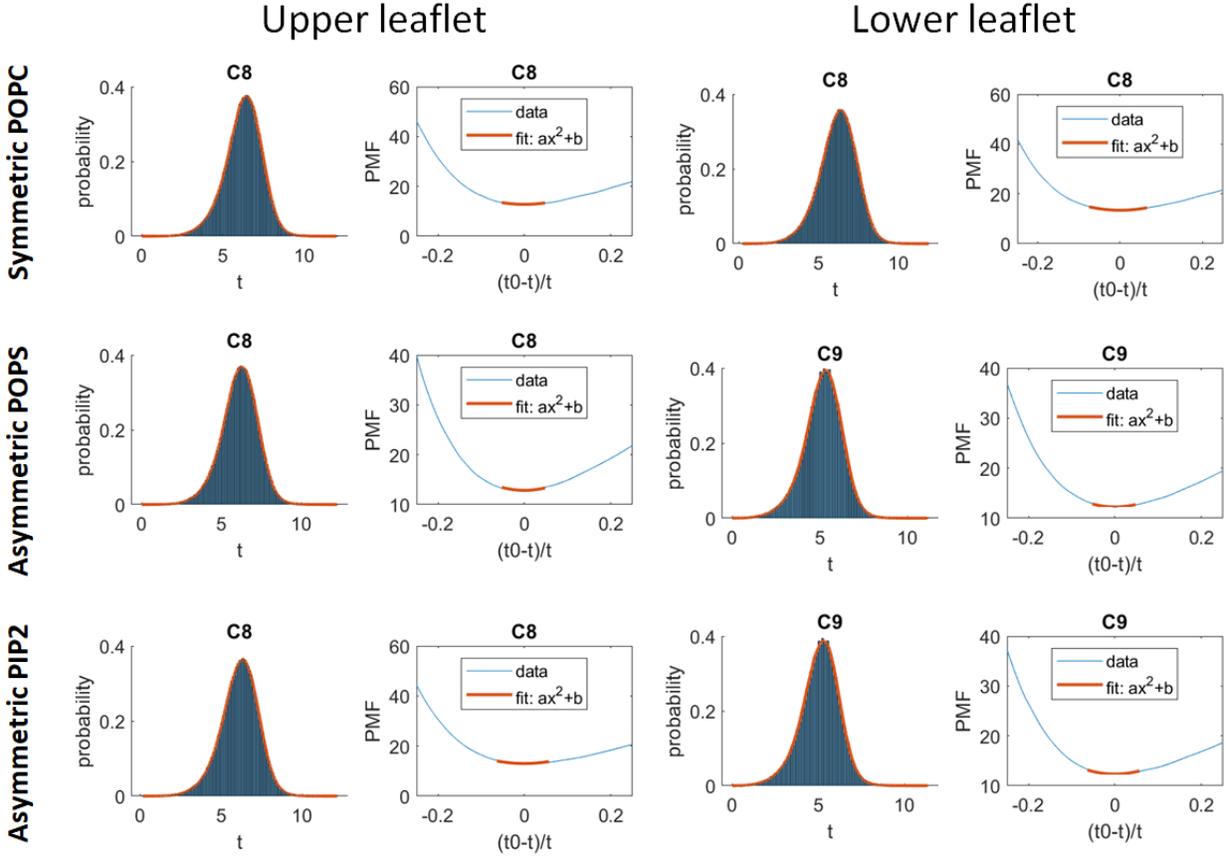

**Figure S3**. Calculation of $K_A^L$ from local thickness fluctuations. For each system, the left plot shows the probability distribution of the relevant leaflet thickness $t$, smoothed using a kernel density. The right plot is $-\frac{2k_BT}{a_0^L}\ln p(\frac{\delta t}{t})$ (i.e., PMF) vs. $\frac{\delta t}{t}$ in Eq.2. The region of thickness within 6% of the mean thickness is fitted to a function of the form $y \sim ax^2 + b$ . $K_A^L$ is obtained from the quadratic coefficient $a$ of the best fit (shown as the thick line segment).

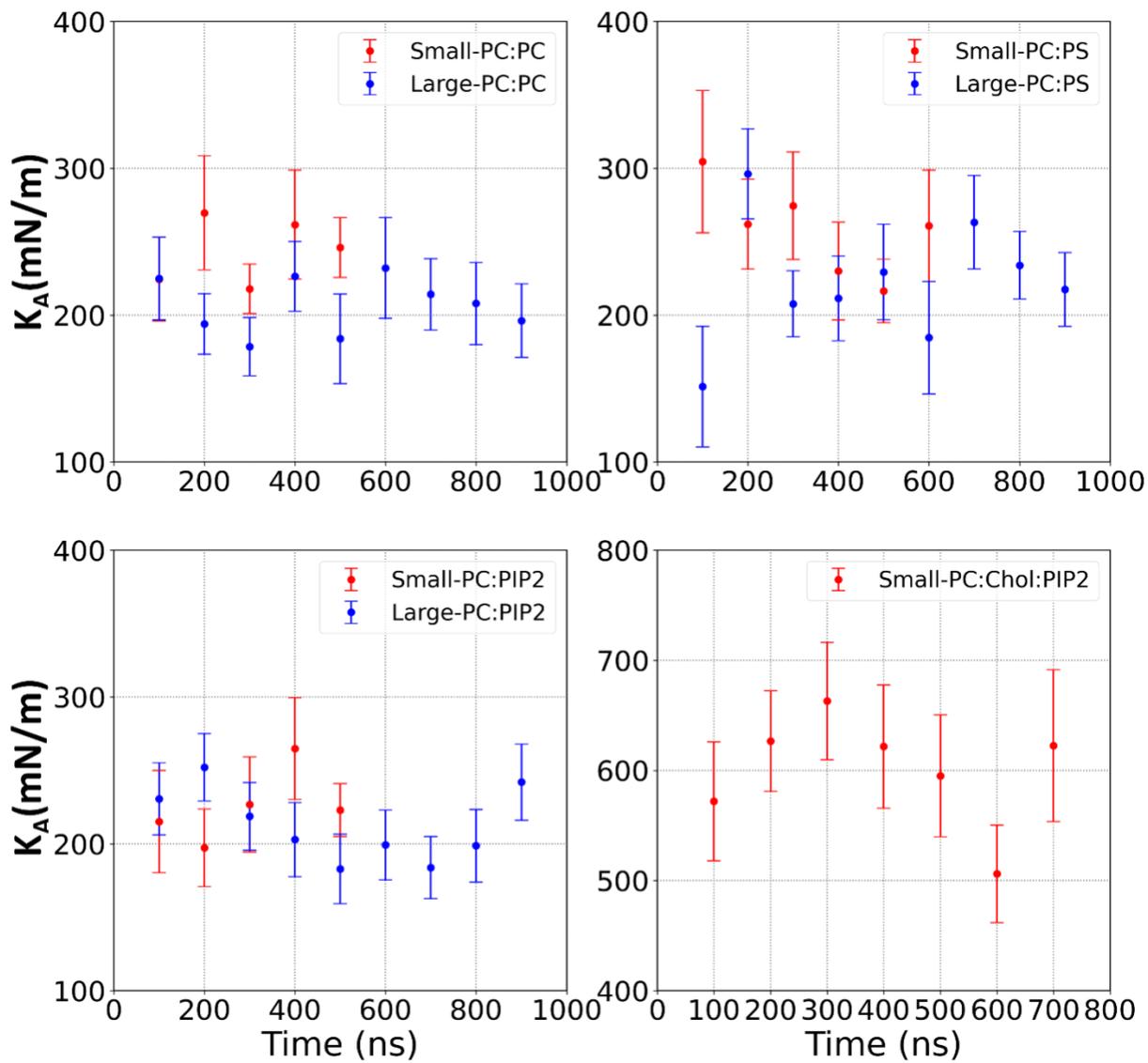

**Figure S4**. Time convergency plot of total bilayer $K_A^L$ from local thickness fluctuations. Large bilayer (in blue) and small bilayer (in red) $K_A^L$ are shown together for comparison. Error bars were calculated from the bootstrapping analysis.

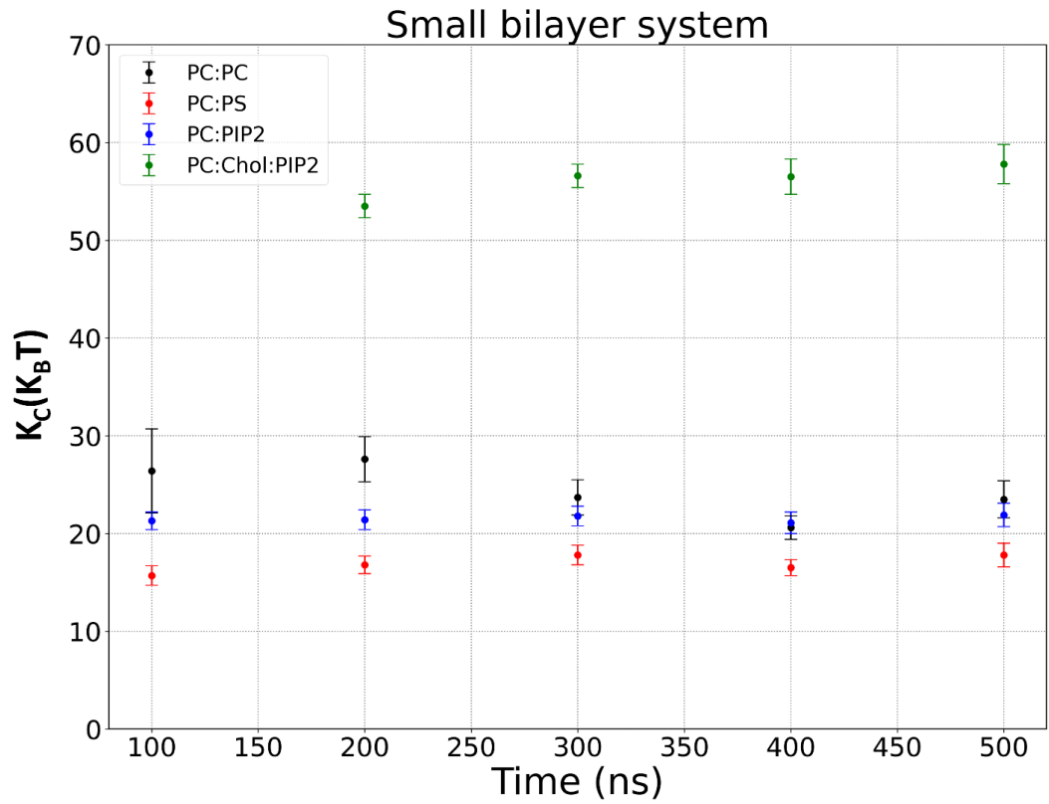

**Figure S5**. Time convergency plot of total bending modulus $K_C$ for small bilayer systems within 500 ns. Error bars were the standard deviations from quadratic fitting using for five fitting ranges (see Methods for detail).

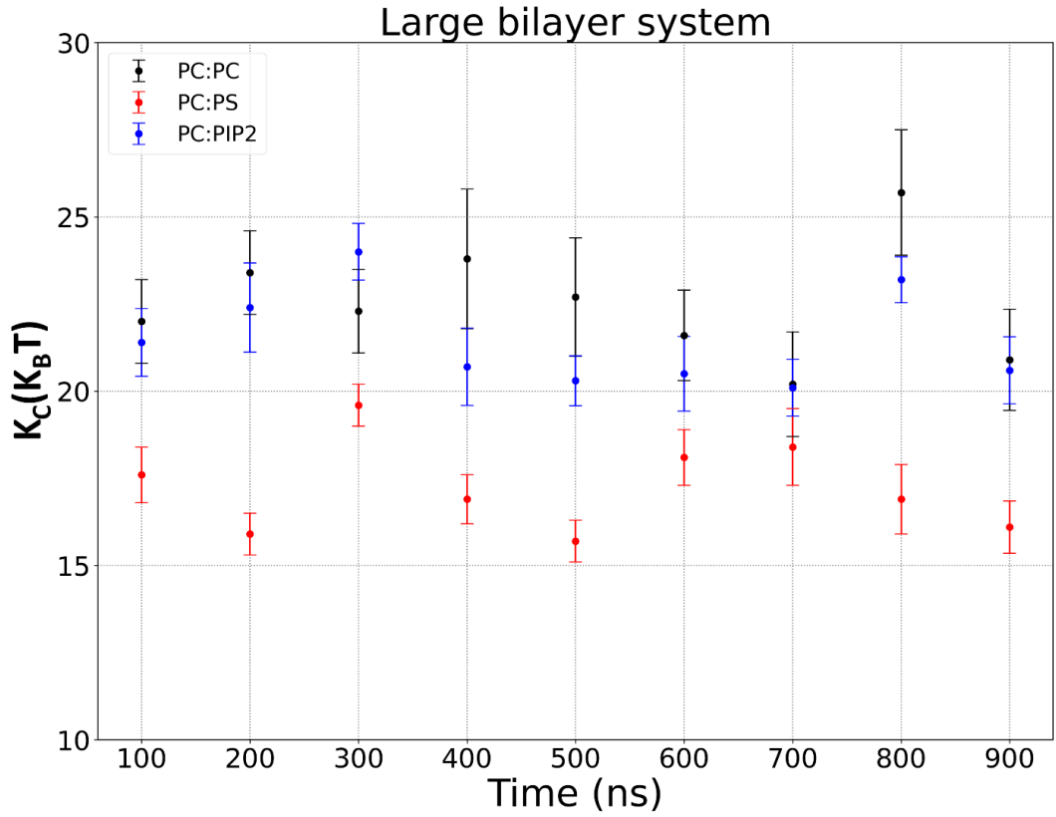

**Figure S6**. Time convergency plot of total bending modulus $K_C$ for large bilayer systems within 900ns. Error bars were the standard deviations from quadratic fitting using for five fitting ranges (see Methods for detail).